\documentclass[preprint2]{aastex}

\usepackage{epsfig}
\usepackage{graphicx}

\begin{document}

\def\hav{\mathop{\mathrm{hav}}\nolimits}
\bibliographystyle{aj} 

\title{Atmospheric Refraction Path Integrals in Ground-Based Interferometry}
\author{Richard J. Mathar}
\affil{Sterrewacht Leiden, Leiden University, Postbus 9513, 2300 RA Leiden, The Netherlands}
\email{mathar@strw.leidenuniv.nl}

\begin{abstract}
The basic effect of the earth's atmospheric refraction on telescope
operation is the reduction of the true zenith angle to the apparent zenith
angle,
associated with prismatic aberrations due to the dispersion in air.
If one attempts coherent superposition of star images in ground-based
interferometry, one is in addition interested in the optical path length
associated with the refracted rays. In a model of a flat earth,
the optical path difference between these is not concerned as the
translational symmetry of the setup means no net effect remains.

Here, I evaluate these interferometric integrals in the more realistic
arrangement of
two telescopes located on the surface of a common earth sphere and point to
a star through an atmosphere which also possesses spherical symmetry.
Some focus is put on working out series expansions in terms of the
small ratio of the baseline over the earth radius, which allows to
bypass some numerics which otherwise is challenged by strong
cancellation effects in building the optical path difference.
\end{abstract}

\section{Pointing and Apparent Star Altitudes}\label{Sec.lateral}
\subsection{Basics: Flat Earth Model}\label{sec.flatearth}
\begin{figure}[h]
\plotone{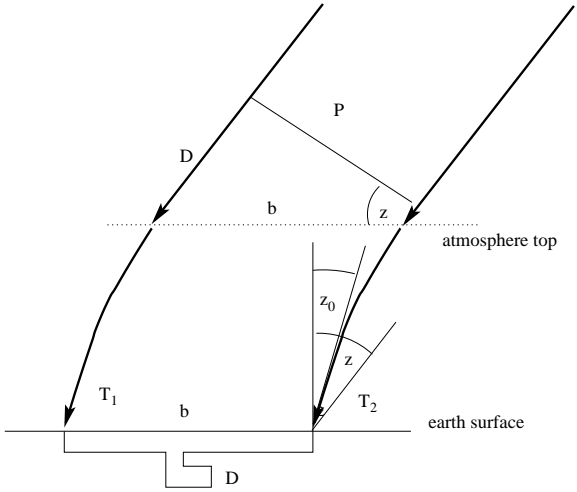}
\caption{
The standard model of delay correction and recovery
for a star at zenith angle $z$
observed by two telescopes that look through horizontal layers
of the stratified atmosphere.
$D=b\sin z=P \tan z$.
}
\label{DelModl.ps}
\end{figure}

The standard model of an interferometric setup
and delay line correction for a star at the true zenith angle $z$ is shown
in Fig.\ \ref{DelModl.ps}:
The optical path difference (OPD) $D$ shows up once in the vacuum above
telescope 1, and is added for telescope 2 on the ground at some local index of
refraction.
The atmosphere is horizontally homogeneous and the earth flat; therefore
no correction is needed for the ray path curvature induced by
any vertical gradient of the index of refraction through the atmosphere, because
these two paths match each other at each height above ground.

The pointing difference $R$ between the apparent and actual altitude
of a star that is induced by the refraction of the earth atmosphere
is in a simple model of a flat earth \citep{Green1985,FilippenkoPASP94}
\begin{equation}
R\approx (n_0-1)\tan z_0,
\label{eq.Rofnflat}
\end{equation}
where $n_0$ is the index of refraction on the ground, $z_0$
the apparent zenith distance on the ground, and
\begin{equation}
R=z-z_0 >0
\end{equation}
obtained in radian.

This implies chromatic effects \citep{Livengood,LanghansAN324,ColavitaPASP116,RoePASP114}
and dependencies on the atmospheric model,
commonly summarized under the label ``transversal atmospheric dispersion'' in
Astronomy.
There is a rainbow effect in Eq.\ (\ref{eq.Rofnflat}): as $n_0$
is a function of the wavelength, $R$ becomes dispersive,
too: between wavelengths of $2$ $\mu$m and  $2.4$ $\mu$m,
we get a difference of $\Delta n_0\approx 1.05\cdot 10^{-7}$ at 2600 m above
sea level,
which translates
into a spectral smear of $\Delta R\approx 22$ mas $\cdot\tan z_0$.
\label{sec.rainbow}

Eq.\ (\ref{eq.Rofnflat})
suggests that the relative error in the star's altitude definition
is close to the relative error in the dielectric function and susceptibility
at the telescope site.
In addition, Snell's law of refraction states that product $n\sin z$ between
the height-dependent index of refraction $n$ and the sine of the angle of
refraction at that height remains constant along the path
\cite[(4.2)]{Green1985}.
Therefore the optical path delay is $D=b\sin z=bn_0\sin z_0$; it changes
as the astronomical object changes position in $z$, or, supposed $b$
is fixed, according to atmospheric parameters accessible at the ground level.
The benefit of this analysis is that both, the pointing correction $R$
in Eq.\ (\ref{eq.Rofnflat}) and the OPD measured on the ground, are functions
of the index of refraction at the telescope site, not functionals
of the entire layered atmosphere.

The theme of this paper are corrections to these statements considering
a non-turbulent atmosphere covering an earth surface of constant, but
non-negligible curvature.
The rest of Sec.\ \ref{Sec.lateral} shortly describes the standard theory
of refraction and defines two different baseline lengths.
Sec.\ \ref{sec.zeni} concentrates on the integral formulation
of the OPD calculation through the atmosphere:
geometries with constrained azimuths suffice to introduce all
relevant concepts; general star positions are then reduced
to the constrained case.

\subsection{Spherical Earth: Geometry}\label{sec.sphergeo}
\begin{figure}[htb]
\plotone{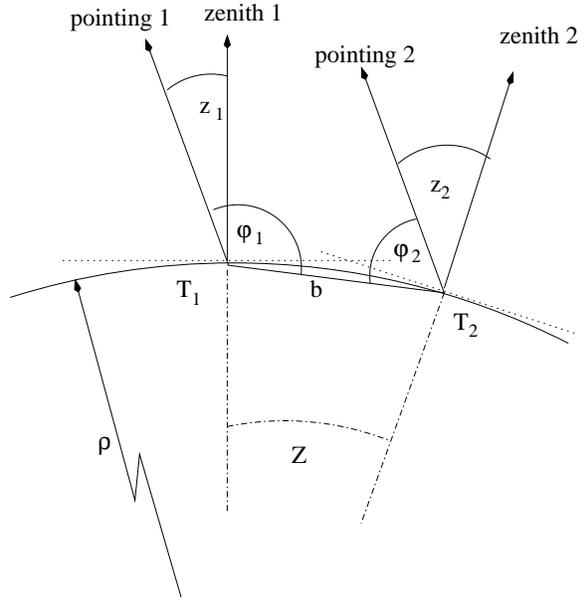}
\caption{
Two telescopes placed on the earth of radius $\rho$ at a distance
$b$ looking at the same star experience local zenith angles that differ
by $Z\approx b/\rho$ rad.
}
\label{PointRho.ps}
\end{figure}

\begin{figure}[h]
\plotone{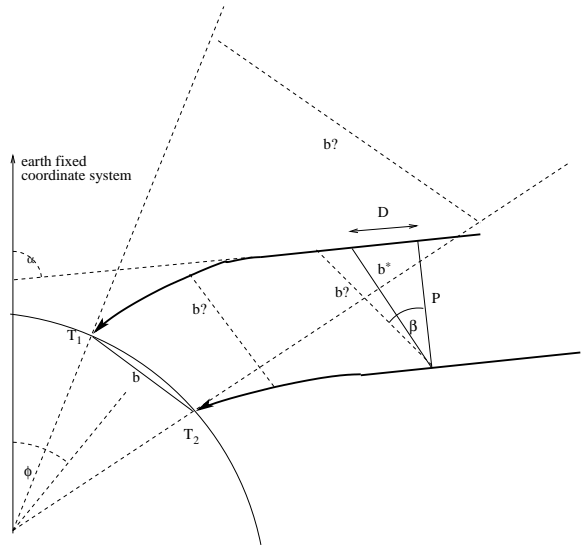}
\caption{
Interferometry with two telescopes located on a non-planar surface
needs to be aware of the vagueness of the definition of a baseline $b$.
}
\label{Base.ps}
\end{figure}

Things are more complicated if we start to look at the more realistic
model of a spherical earth. A telescope distance of $b=100$ m
on an earth of radius $\rho=6368$ km leads to a pointing mismatch  of purely
geometric origin
of about $b/\rho\approx 3.2$ arcsec (Fig.\ \ref{PointRho.ps}).
The baseline $b$ is the distance between the telescope locations on the
earth, the length of the straight secant drawn in
Figs.\ \ref{PointRho.ps} and \ref{Base.ps},
\begin{eqnarray}
b&=&\rho\sqrt{2(1-\cos Z)} \label{eq.bofZ} \\
&=&2\rho\sin\frac{Z}{2} \\
&\approx& \rho\left(Z-\frac{Z^3}{24}+\frac{Z^5}{1920}\ldots \right).
\label{eq.bforz}
\end{eqnarray}
The inversion of this series reads
\begin{equation}
Z\approx \frac{b}{\rho}
+\frac{1}{24}\left(\frac{b}{\rho}\right)^3
+\frac{3}{640}\left(\frac{b}{\rho}\right)^5\ldots.
\label{eq.zforb}
\end{equation}
The angle approximation $Z\approx b/\rho$ is an estimate to $b/\rho =2\sin (Z/2)$,
a limit of a baseline so short that it does not matter whether it is 
measured along a straight line (as drawn in Fig.\ \ref{PointRho.ps})
or along the circular perimeter.
The relative error in this approximation
is $\approx Z^2/24$, or $\approx 10^{-11}$ for the example of
$b=100$ m.

Pointing/guiding is a functionality of the individual telescopes: the existence
of a nonzero pointing difference is absorbed in the telescope
operation, and any delay originating from there is to first order recovered
in the tracking.
We hereby explicitly discard any ``separation'' term of Gubler and Tytler
eminent from their consideration of single telescopes
\citep{GublerPASP110}, and acknowledge that both telescopes
are ``mispointing'' at the same time.

This cross-eyed geometry indirectly transforms into a contribution to
the delay that may be understood---and calculated to lowest order---on
the basis of what is said in Sec.\ \ref{sec.flatearth}, but has no
parallel with single telescopes as long as their diameters are much
smaller than the earth radius: The example of $3.2$ arcsec
from above translates into an additional angle of
refraction of
$\Delta R\approx \Delta z_0(1+\tan^2z_0)(n_0-1)>\Delta z_0(n_0-1)\approx
Z (n_0-1)\approx 650$ $\mu$as at 2600 m above sea level,
which is the first derivative of Eq.\ (\ref{eq.Rofnflat}) w.r.t.\ $z_0$.
Dropping the term $\tan^2z_0$ here means this is a lower estimate
in the limit of stars at the zenith.

The effect on the delay could be understood in the
``standard model of delay line correction,'' where the two rays of
the star that will eventually hit the two telescopes
``generate'' a phase difference in the vacuum (undone later on
in the delay line tunnel) as they hit the top layer of the earth
atmosphere with a path difference $D$
(Fig.\ \ref{DelModl.ps}). The curvature correction means
that this top layer ``bends back'' a bit more for the telescope further away
from the star, which slightly increases the angle of incidence on the
atmosphere.

Fig.\ \ref{Base.ps} shows that the only obvious definition of the
baseline length $b$ is on earth.
Three alternatives have been marked with question 
marks in the figure:
\begin{enumerate}
\item
Simple outward projection along some
``common'' zenith direction of both telescopes is futile, because
it would only scale $b$ with some arbitrary distance from
the earth center.
\item
Some sliding definition along the curved
rays fails because there are no fix points or right angles for reference.
\item
The baseline cannot be rigorously defined
far outside the atmosphere. Actually only the projection $P$
of the two telescopes remains well defined as the rays can be considered
parallel, and so remains some direction angle $\alpha$ relative to some
geostationary coordinate system.
One could try to use the b? in Fig.\ \ref{Base.ps} as some line parallel to
the earth-based $b$ anchored at $P$, but needs to stay aware of the
fact that this replaces the zenith angle $z$ in Fig.\  (\ref{DelModl.ps})
by the angle $\beta=\alpha-\phi$, where $\alpha$ is the polar coordinate
of the star and $\phi$ the mean polar coordinate of the telescopes
in the geocentric coordinate system.
\end{enumerate}
In a formal way, I define an effective baseline length $b^*$
above the atmosphere via
\begin{equation}
b^{*2} \equiv P^2+D^2
\label{eq.beff}
\end{equation}
which turns out to be longer than $b$ because the earth's atmosphere represents
a gradient lens \citep{HuiAJ572}. This effective baseline is a function of the baseline $b$
and of the star zenith angle, and a functional of the atmospheric refraction
$n(r)$.

\subsection{Spherical Earth: Atmospheric Layers}\label{sec.spherlay}
The delay line modification described in the previous paragraph is
of purely geometrical nature and is closely related to the discussion
of differences between two light rays, but there are more pointing
``corrections'' that already show up in the single ray case, as discussed
next.

Eq.\ (\ref{eq.Rofnflat}) is based on the assumption that the gradient
of the refractive index $n$ along the ray path is parallel to a global, constant
zenith vector. As the gradient and the zenith vector change in direction
along the path through a spherically symmetric atmosphere, and as we assume
that $n$ becomes a function of the radial distance to the spherical surface of
the earth, the equation becomes more accurately
\cite[(4.19)]{Green1985}\citep{ThomasJHUAPL17,AuerAJ119,NenerJOSA20,NoerdlingerJPRS54,Tannousarxiv01}
\begin{equation}
R=\rho n_0\sin z_0\int_1^{n_0}\frac{dn}{n(r^2n^2-\rho ^2n_0^2\sin^2z_0)^{1/2}},
\label{eq.RofnInt}
\end{equation}
an integral over the refractive index, to start above the atmosphere
($n=1$) and to end at the telescope position ($n=n_0$), and where
$r\ge \rho$ is the distance to the earth center.
Its Taylor expansion is
commonly written as an expansion in powers of $\tan z_0$ \citep{StonePASP108},
the observed quantity,
\begin{eqnarray}
R&=&\rho n_0\tan z_0\int_1^{n_0}\frac{dn}{n^2r} \nonumber \\
&&+\frac{\rho n_0\tan^3 z_0}{2} \int_1^{n_0}\frac{\rho^2n_0^2-r^2n^2}{n^4r^3}dn \nonumber \\
&& +\frac{3\rho n_0\tan^5 z_0}{8} \int_1^{n_0}
\frac{(\rho^2n_0^2-r^2n^2)^2}{n^6r^5}dn
+\ldots.
\end{eqnarray}
The major new aspect is that the pointing direction becomes a functional
of the height spectrum of the index $n$, and the angle of arrival becomes
site-dependent \citep{ConanJOSA17}.

\begin{figure}[h]
\plotone{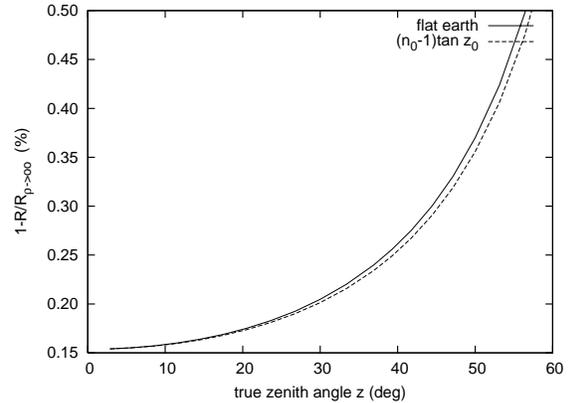}
\caption{
Solid line: The relative error in the refraction angle $R$, in percent, made
by switching from the spherical
model, Eq.\ (\ref{eq.RofnInt}), to the limit of the
flat earth, Eq.\ (\ref{eq.Rflat}). Dashed line: The relative error by
switching from the spherical model to the approximation (\ref{eq.Rofnflat}).
The {\em absolute} errors approach 0 as $z\to 0$.
}
\label{Rdiff.ps}
\end{figure}

Accurate modeling of $n(r)$ is not within the scope of this treatise here
and unambitiously discussed in App.\ \ref{sec.nofr}\@.
Tables and graphs to follow use an exponential depletion of the susceptibility
\begin{equation}
\chi=\epsilon-1=n^2-1
\label{eq.chiofn}
\end{equation}
to the vacuum of the universe with a scale height $K$,
\begin{equation}
\chi(r)=\chi_0e^{-(r-\rho )/K}.
\label{eq.chiofr}
\end{equation}
The explicit parameters are
\begin{equation}
\chi_0=4\cdot 10^{-4},\quad \rho=6380 \mathrm{ km},\quad K=10 \mathrm{ km},
\label{eq.chiofrPar}
\end{equation}
unless otherwise noted, representing a prototypical K-band value at 2600 m
above sea level \citep{MatharAO43}.
Within this model, Fig.\ \ref{Rdiff.ps} shows the change in $R$ introduced by
switching from the flat earth model,
\begin{equation}
R \stackrel{\rho\to \infty}{\longrightarrow}
n_0\sin z_0\int_1^{n_0}\frac{dn}{n(n^2-n_0^2\sin^2z_0)^{1/2}},\label{eq.Rflat}
\end{equation}
to the spherical model of the atmosphere.

\section{Accumulated Optical Path Lengths for Spherical Geometry}\label{sec.zeni}
\subsection{Planar, Overhead Geometry}\label{sec.zeni2D}
\subsubsection{Single Star}
\begin{figure}[hbt]
\plotone{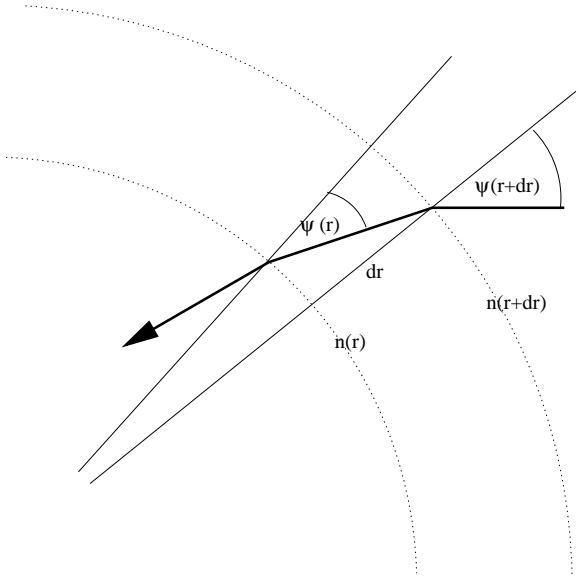}
\caption{The integrals (\ref{eq.RofnInt}) and (\ref{eq.Daux}) accumulate
the change in the geocentric zenith angle $\psi(r)$ and the optical
path length $n/\cos\psi$, respectively, along the light path, applying
Snell's law $r n \sin\psi(r)=$ const at each differential layer.
}
\label{Psi.ps}
\end{figure}

\begin{figure}[hbt]
\plotone{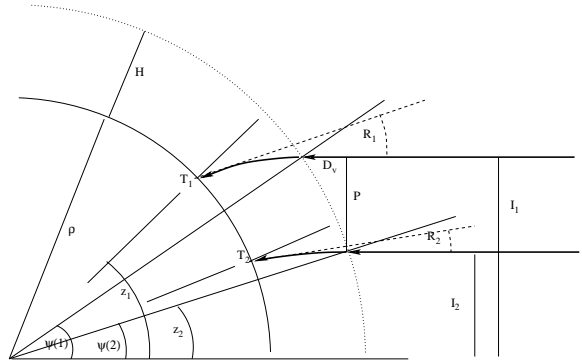}
\caption{
Sketch of the impact parameters $I_i$, the auxiliary geocentric
zenith angles $\psi^{(i)}_H$, the true zenith angles
$z_i$, the angles of refraction
$R_i$, and the projected baseline $P$. The two light rays enter
from the right, hit the upper atmosphere at a height $H$ where indicated by the
dotted quarter-circle, and eventually the telescopes $T_i$.
}
\label{DelModl2.ps}
\end{figure}

The optical path length along the curved ray trajectory is the line integral
of the refractive index over the geometric path, similar to
Eq.\ (\ref{eq.RofnInt})
\begin{equation}
L=\int_{n=1}^{n_0}n \frac{dr}{\cos \psi},
\label{eq.Daux}
\end{equation}
where $dr/\cos \psi$ is the length of the diagonal path element in
Fig.\ \ref{Psi.ps}.
$\psi$ the local zenith angle of the star, and the $r$ the distance
from the earth center
\cite[Fig. 4.4]{Green1985}.
This quantity includes the geometric path length up to the star and is
infinite in our applications. 
We assume a maximum height $H$ of the atmosphere, $n_{|r>\rho+H}=1$, with
the option to look at the limit $H\to\infty$ if the density has no
clear ceiling, like in the models of Eq.\ (\ref{eq.chiofr}).
We define the impact parameters $I_1$ and $I_2$ of the rays,
$0\le I_2\le I_1\le \rho$, which were their
smallest distances to the earth center if they would pass by along geometric
straight lines without any diffraction---as used in atomic collision theory.
\begin{equation}
P=I_1-I_2
\label{eq.PofIs}
\end{equation}
is the projected baseline, measured above the atmosphere.
Sec.\ \ref{sec.zeni2D} considers the case
in which the star,
the two telescopes (the baseline), and the earth center
are coplanar: Fig.\ \ref{DelModl2.ps}\@.
Comments on the general configuration of an unconstrained star azimuth follow in
Sec.\ \ref{sec.zeni3D}.

The difference between the integrals
Eq.\ (\ref{eq.Daux})
\begin{eqnarray}
&&D=L_1-L_2\\
&&\!=\!\int\limits_{r=\rho}^{\sqrt{(\rho+H)^2+I_1^2-I_2^2}}
\!\!\!\!
\frac{n\, dr}{\cos \psi^{(1)}}
-\int\limits_{r=\rho}^{\rho+H}\!\frac{n\, dr}{\cos \psi^{(2)}},
\end{eqnarray}
is the OPD for two telescopes at a common baseline $b$, with different
apparent and different true zenith angles (individual pointing), but looking
at the same star.
The additional straight line segment for the ray to telescope 1 through vacuum
before it reaches the altitude $H$ is
\begin{eqnarray}
D_v &\equiv& \sqrt{(\rho+H)^2-I_2^2}-\sqrt{(\rho+H)^2-I_1^2} \nonumber \\
&=& (\rho+H)[\cos \psi_H^{(2)}-\cos \psi_H^{(1)}].
\label{eq.Dv}
\end{eqnarray}
The Taylor series of this term up to third order in $P$ is
\begin{equation}
D_v\approx P\tan \psi_H^{(2)}+\frac{1}{2I_2}
\frac{\tan \psi_H^{(2)}}{\cos^2 \psi_H^{(2)}}P^2
+\frac{1}{2I_2^2}
\frac{\tan^3 \psi_H^{(2)}}{\cos^2 \psi_H^{(2)}}P^3.
\label{eq.taylDv}
\end{equation}
We separate this
piece from the integral for the telescope 1,
\begin{equation}
D =D_v+\int_{r=\rho}^{\rho+H} n\Big[\frac{1}{\cos\psi^{(1)}} \nonumber\\
-\frac{1}{\cos\psi^{(2)}}\Big]dr,
\end{equation}
where the major difference w.r.t.\ Fig.\ \ref{DelModl.ps} is that this
integral does not vanish, because the atmosphere is now hit at two different
angles $\psi_H^{(1)}\neq \psi_H^{(2)}$.
The integral would also not vanish, if the factor $n$ would be dropped to
deduce the {\em geometric} path difference of the curved beams.

The constance of the product $rn\sin\psi$ along each curved trajectory
\cite[(4.16)]{Green1985},
\begin{eqnarray}
rn\sin\psi^{(1)} &=&(\rho+H)\sin \psi_H^{(1)}=I_1 ; \\
rn\sin\psi^{(2)} &=&(\rho+H)\sin \psi_H^{(2)}=I_2 ,
\label{eq.Snell}
\end{eqnarray}
is inserted into the previous equation,
\begin{eqnarray}
D =D_v&+&\int\limits_{r=\rho}^{\rho+H} rn^2\Big[
\frac{1}{\sqrt{r^2n^2-(\rho+H)^2\sin^2\psi_H^{(1)}}}\nonumber\\
&&-\frac{1}{\sqrt{r^2n^2-(\rho+H)^2\sin^2\psi_H^{(2)}}}\Big]dr.
\label{eq.DIntf}
\end{eqnarray}

The term in square brackets allows another Taylor expansion
\begin{eqnarray}
&&\frac{1}{\sqrt{r^2n^2-I_1^2}}-\frac{1}{\sqrt{r^2n^2-I_2^2}}
\approx \nonumber \\
&& \frac{I_2}{(r^2n^2-I_2^2)^{3/2}}P
+\frac{1}{2}\frac{r^2n^2+2I_2^2}{(r^2n^2-I_2^2)^{5/2}}P^2 \nonumber \\
&&+\frac{1}{2}\frac{3r^2n^2+2I_2^2}{(r^2n^2-I_2^2)^{7/2}}I_2P^3.
\label{eq.TaylIntf}
\end{eqnarray}
All correction terms of the spherical geometry in
Eqs.\ (\ref{eq.taylDv})--(\ref{eq.TaylIntf}) have a positive sign.
The contribution of the term of $O(P^2)$ amounts to $\approx 2$  mm,
and of the term of $O(P^3)$ to $\approx 60$ nm as the zenith angle approaches
$60$ deg at $b=100$ m.
A consistency check of Eq.\ (\ref{eq.TaylIntf})  is that its contribution of the
first, linear Taylor order to the integral in Eq.\ (\ref{eq.DIntf})  becomes
\begin{eqnarray}
&&
\int_{r=\rho}^{\rho+H} rn^2\frac{I_2P}{(r^2n^2-I_2^2)^{3/2}}dr \nonumber\\
&&\stackrel{n\to 1}{\longrightarrow}-P\tan \psi_H^{(2)}+P\frac{I_2}{\sqrt{\rho^2-I_2^2}}
\end{eqnarray}
in the vacuum limit, such that the term $P \tan\psi_H^{(2)}$ cancels
the first term in Eq.\ (\ref{eq.taylDv}), and only the term $P\tan z$ depending
on the true topocentric zenith angle remains, equivalent to
Fig.\ \ref{DelModl.ps}.

\begin{figure}[h]
\plotone{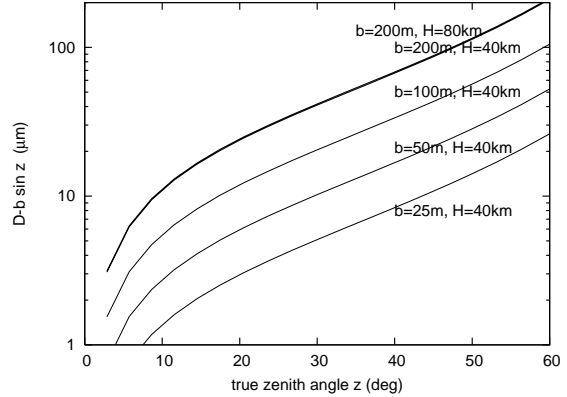}
\caption{
The difference between the full integral (\ref{eq.DIntf}) of the
delay $D$ and the approximation $D\approx b\sin \bar z$, which one would
derive from Fig.\ \ref{DelModl.ps} using the mean true zenith
angle $\bar z \equiv (z_1+z_2)/2$, for baselines between
25 and 200 m. A convergence test to the limit $H\to\infty$
is indicated for $b=200$ m.
}
\label{Dds.ps}
\end{figure}

The influence of the spherical geometry on the atmospheric path length
difference is demonstrated in Fig.\ \ref{Dds.ps}.

\subsubsection{Astrometry (two true zenith distances)}
Up to here, an accurate computation of the delay $D$ by integration over the
atmosphere layers is not competitive against measuring the equivalent value
on the ground (Fig.\ \ref{DelModl.ps}), since the gas densities on the
ground along the beam path---input to calculation of the refractive
indexes---are accessible to sensors and much better known 
than the remote, high-flying layers of air.
As we turn to the astrometric task of completing the right-angled triangle
formed by $D$, $P$ and $b^*$ in Fig.\ \ref{Base.ps}---with the aim
of precise determination of either of the base angles at $b^*$---,
the ``baseline calibration'' emerges as an additional focus.
This means determination of $b^*$, the image of $b$. The following sections
consider the atmospheric lensing correction $b^*-b$ to a, in principle,
rock-solid and accessible ground baseline $b$.
The computational strategy is to derive $P$ and $D$, then to use
Eq.\ (\ref{eq.beff}).
Within this framework, the geometric baseline $b$ is defined joining
the ``ends of the paths through the atmosphere,'' and is assumed to
be the same
vector for both stars in the case of astrometry. We do not ask the
question whether the corresponding terminal points $T_1$ and $T_2$ are that
well defined for real telescope optics with chromatic, azimuth
dependent foci and trusses that bent under the load of their own weight
or the wind.

\begin{figure}[hbt]
\plotone{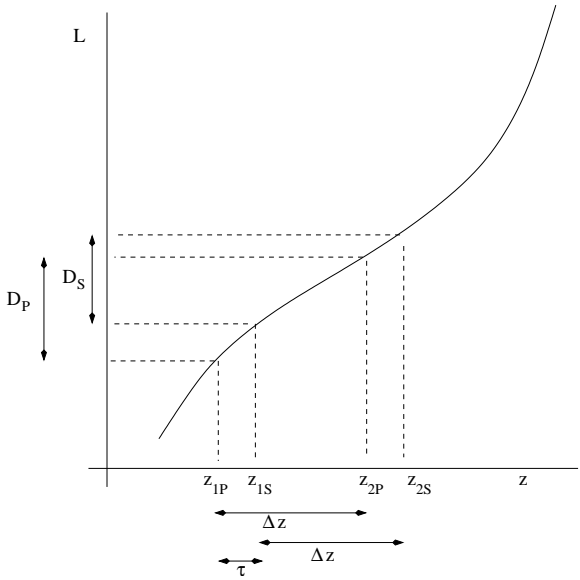}
\caption{A difference in the delays $D_\textrm{P}$ and $D_{\textrm{S}}$
measured for the primary star and the secondary star is caused by
the nonlinearity of the path length $L$, Eq.\ (\ref{eq.Daux}), as a function
of the true zenith position.
}
\label{Lofz.ps}
\end{figure}

The formal solution to the problem of tracing four beams
(two stars, separated by an angle $\tau$ and labeled $P$ for ``primary''
and $S$ for ``secondary,'' to two telescopes) for a given $Z$ in
Eq.\ (\ref{eq.zforb}) is then given by computation of $D_\textrm{P}$ for the
primary at some $Z$ (some baseline $b$, see Eq.\ (\ref{eq.bforz})),
computation of $D_\textrm{S}$ for the secondary at the
same $\Delta z=Z=z_{1\textrm{S}}-z_{2\textrm{S}}
=z_{1\textrm{P}}-z_{2\textrm{P}}$
(since the secondary must be caught by the same two telescopes)
but slightly different true zenith angles
$z_{1\textrm{S}}= z_{1\textrm{P}}+\tau$, 
$z_{2\textrm{S}}= z_{2\textrm{P}}+\tau$.
It should be noted that the earth-bound baseline is {\em not\/} tilted here---
there is no lifting of one telescope and sinking of the other
to acquire the secondary---, and $b$ refers
to the geometrical distance between two foci that define a
common reference for both stars.
Any residual effects
of the spherical atmosphere and/or atmospheric layering are then
caused by the second derivative of $L(z)$ (Fig.\ \ref{Lofz.ps}). 

From (\ref{eq.DIntf}) with (\ref{eq.Snell}), omitting a constant $\rho+H$,
\begin{eqnarray}
L=&& -\sqrt{(\rho+H)^2-\rho^2n_0^2\sin^2 z_0} \nonumber \\
&&+\int\limits_{r=\rho}^{\rho+H}\!\! 
\frac{rn^2}{(r^2n^2-\rho^2n_0^2\sin^2 z_0)^{1/2}}dr, \label{eq.LInt}
\end{eqnarray}
where we insert $z_0=z-R$.
For a spherical earth without atmosphere, the vacuum limit,
\begin{equation}
R\stackrel{n\to 1}{\longrightarrow} 0,\quad n_0\longrightarrow 1,\quad
\end{equation}
\begin{eqnarray}
L& \longrightarrow& \rho(1-\cos z)\nonumber \\
&=& \frac{\rho}{2}\tan^2z
-\frac{3}{8}\rho\tan^4z+\frac{5}{16}\rho\tan^6z\ldots,
\end{eqnarray}
\begin{equation}
D\longrightarrow \rho(\cos z_2-\cos z_1).
\label{eq.Dvac}
\end{equation}
This suggests we define an effective true zenith angle $\bar z$ via
\begin{equation}
D\equiv b\sin \bar z=P\tan \bar z,
\end{equation}
and use Eqs.\ (\ref{eq.bforz}) and (\ref{eq.Dvac}) to prove that this equals
the mean,
\begin{equation}
\bar z=\frac{z_1+ z_2}{2}.
\label{eq.zbar}
\end{equation}
Since the astrometric signature is hidden in the second derivative of $L(z)$,
the computationally most appealing expansion of $R$ and $L$ is
a Taylor series around some reference value of $z$:
\begin{equation}
R_{|z+x}\equiv \xi_0+\xi_1x+\xi_2x^2+\xi_3x^3+\ldots, \qquad \xi_0 = R_{|z}.
\label{eq.rtayl}
\end{equation}
We define refractivity integrals
as a short-cut to the notation, covering
Eq.\ (\ref{eq.RofnInt}) as a special case:
\begin{eqnarray}
R_j& \equiv& I^{2j+1}\int_1^{n_0}\frac{dn}{n(r^2n^2-I^2)^{j+1/2}}, \label{eq.RofnInt2}\\
I& \equiv& \rho n_0\sin z_0, \qquad j=0,1,2,\ldots
\end{eqnarray}
A stable numerical scheme for these integrals is proposed
in App.\ \ref{sec.Rnum}\@.
Insertion of the series (\ref{eq.rtayl}) into the l.h.s.\ of
(\ref{eq.RofnInt}) and into the arguments $z_0=z-R$ of the sines at the r.h.s.\ 
yields the expansion coefficients
\begin{eqnarray}
\xi_0&=&R_0, \\
\xi_1 &=& \frac{R_0+R_1}{\tan z_0}/ \left( 1+\frac{R_0+R_1}{\tan z_0} \right), \\
\xi_2&=&\frac{(\xi_1-1)^3}{2}
    \left[R_0+R_1-3\frac{R_1+R_2}{\tan^2z_0}\right] \nonumber \\
&=& \frac{[3R_{12}-R_{01}\tan^2 z_0]\tan z_0}{2 \hat R_{01}^3}, \\
\xi_3 &=&  -(\xi_1-1)^2
    \Big[
     -(R_0+R_1)\xi_2
     +3\frac{R_1+R_2}{\tan^2z_0}\xi_2 \nonumber \\
     &&\quad +\frac{(\xi_1-1)^2}{\tan z_0}
     \Big\{
        \frac{1}{6}R_0
        +\frac{5}{3}R_1-3\frac{R_2}{\tan^2z_0} \nonumber \\
     &&\quad   -\frac{1}{2}\frac{R_1}{\tan^2 z_0}
        -\frac{5}{2}\frac{R_3}{\tan^2 z_0}+\frac{3}{2}R_2
	\Big\}
     \Big] \\
&=& -\frac{\tan z_0}{6\hat R_{01}^5}\Big[
15(R_1-R_3)\hat R_{01} \nonumber \\
&& +9R_{12}(3R_{12}-2\hat R_{01}-\bar R_{01}\tan^2 z_0) \nonumber \\
&& +R_{01}(\hat R_{01}+3R_{01}\tan^2z_0)\tan^2z_0
\Big] ,
\end{eqnarray}
with the doubly-indexed shorthands
\begin{eqnarray}
R_{ij}&\equiv& R_i+R_j,\\
\hat R_{ij}&\equiv& R_i+R_j+\tan z_0,\\
\bar R_{ij}&\equiv& R_i+R_j-\tan z_0. \label{eq.Renddef}
\end{eqnarray}
In Eqs.\ (\ref{eq.RofnInt2})--(\ref{eq.Renddef}) and App.\ \ref{sec.Rnum},
the subscripts of $R$ are the exponential $j$ of the
definition (\ref{eq.RofnInt2}); elsewhere they indicate the
telescope number/site.
\begin{figure}[hbt]
\plotone{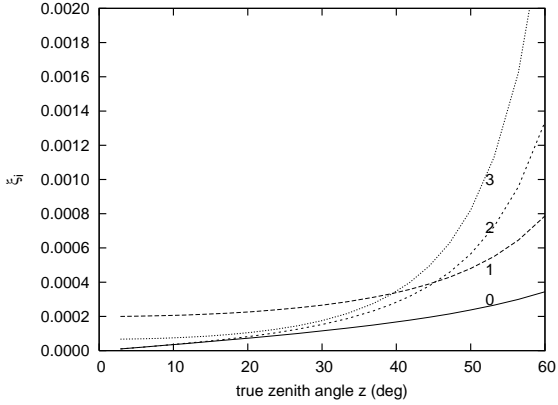}
\caption{
The expansion coefficients $\xi_j$ of Eq.\ (\ref{eq.rtayl}) as a function
of zenith angle $z$ for $j=0,\ldots,3$ and the atmospheric model
(\ref{eq.chiofr})--(\ref{eq.chiofrPar}).
$R$ is an odd function of $z$, so the even-indexed $\xi_j$ approach 0 for
$z\to 0$. $\xi_1$ approaches $n_0-1$, see Eq.\ (\ref{eq.Rofnflat}).}
\label{Rtayl.ps}
\end{figure}
$x$ in Eq.\ (\ref{eq.rtayl}) is of the order of $b/(2\rho)$ if the reference
azimuth $z$
is chosen close to the middle between the telescopes, and therefore
not larger than $1.6\cdot 10^{-5}$ rad for $b<200$ m. Because the  $\xi_j$ are 
approximately of the same magnitude (Fig.\ \ref{Rtayl.ps}), collecting the
terms up to $j=3$ ought establish a relative accuracy of
$\approx 5\cdot 10^{-14}$ in the angle of refraction.

The expansion
\begin{equation}
L_{|z+x} = L_{|z}+\sum_{i=1,2,3,\ldots} l_i x^i, \label{eq.Lofl}
\end{equation}
proceeds via insertion of Eq.\ (\ref{eq.rtayl}) into the sines of
Eq.\ (\ref{eq.LInt}), and employs an auxiliary set of integrals
\begin{eqnarray}
v_i\equiv  I^{2i}\Big[ && 
\frac{1}{[(\rho +H)^2-I^2]^{i-1/2}} \nonumber \\
&&+(2i-1)
\int_{r=\rho}^{\rho+H} \frac{rn^2}{(r^2n^2-I^2)^{i+1/2}}dr\Big] \nonumber \\
&& \stackrel{n\to 1, I\to \rho\sin z}{\longrightarrow} \rho\tan^{2i}z\cos z
.
\end{eqnarray}
\begin{eqnarray}
l_1 &=& \frac{1-\xi_1}{\tan z_0}v_1
\stackrel{n\to 1, I\to \rho\sin z}{\longrightarrow} \rho\sin z,
\\
l_2 &=& -\frac{1}{2}v_1
\left[(1-\frac{1}{\tan^2 z_0})(1-\xi_1)^2+\frac{2\xi_2}{\tan z_0}\right] \nonumber \\
&& +\frac{1}{2}v_2\frac{(1-\xi_1)^2}{\tan^2 z_0} \\
&& \stackrel{n\to 1, I\to \rho\sin z}{\longrightarrow} \frac{1}{2}\rho\cos z, \label{eq.l2vac}
\\
l_3 &=& 
 \frac{1}{3}v_1(1-\xi_1)
\left[-2\frac{(1-\xi_1)^2}{\tan z_0}+3\xi_2(1-\frac{1}{\tan^2 z_0})\right] \nonumber \\
&& -\frac{1}{2}v_2\frac{1-\xi_1}{\tan z_0}
 \left[(1-\frac{1}{\tan^2 z_0})(1-\xi_1)^2+\frac{2\xi_2}{\tan z_0}\right] \nonumber \\
&&\qquad +\frac{1}{2}v_3\frac{(1-\xi_1)^3}{\tan^3 z_0}
\stackrel{n\to 1, I\to \rho\sin z}{\longrightarrow} -\frac{1}{6}\rho\sin z, \nonumber
\\
l_4 && 
\stackrel{n\to 1, I\to \rho\sin z}{\longrightarrow} -\frac{1}{24}\rho\cos z .
\end{eqnarray}
\begin{figure}[hbt]
\plotone{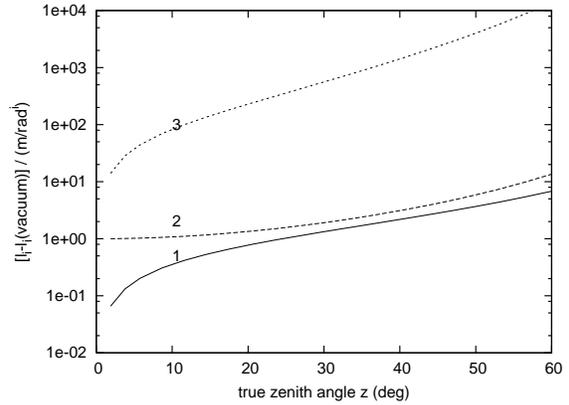}
\caption{
The differences between the Taylor expansion coefficients $l_j$ of
Eq.\ (\ref{eq.Lofl}) and their vacuum values as a function
of zenith angle $z$ for $j=1,\ldots,3$, parametrized through the exponential
model (\ref{eq.chiofr})--(\ref{eq.chiofrPar}).
$L$ is an even function of $z$, so the odd-indexed $l_j$ approach 0 for
$z\to 0$.  The vacuum limits, Eq.\ (\ref{eq.Dvac}), are $l_1\to \rho\sin z$,
$l_2\to \rho(\cos z)/2$, $l_3\to -\rho(\sin z)/6$.
}
\label{Ltayl.ps}
\end{figure}
If the $l_i$ are calculated at the mean position $\bar z=(z_1+z_2)/2$,
the delay for a single sky position becomes
\begin{eqnarray}
D &\approx&
l_1\left(z_1-z_2\right)
+l_2\left(z_1^2-z_2^2\right) \nonumber \\
&& +l_3\left(z_1^3-z_2^3\right)
+l_4\left(z_1^4-z_2^4\right)+\cdots \\
&\approx&
l_1\Delta z
+l_3 (\Delta z)^3/4 + l_5(\Delta z)^5/16 \nonumber \\
&& +l_7(\Delta z)^7/64+\cdots \label{eq.Dofl}
\end{eqnarray}
If the $l_i$ are calculated at the mean position $\bar z=(z_{1\textrm{P}}+z_{2\textrm{P}})/2$,
the differential delay is to lowest orders in $\tau$
\begin{eqnarray}
\!\!\!&&\!\!\!\!\!D_\textrm{S}-D_\textrm{P} \nonumber \\
\!\!\!&&\!\!\!\!\!=\Delta z\tau \Big[
2 l_2
+3\tau l_3
+\left((\Delta z)^2+4\tau^2\right)l_4 \nonumber \\
&& +\left(\frac{5}{2}(\Delta z)^2\tau+5\tau^3\right)l_5 \nonumber \\
&&+ \left(\frac{3}{8}(\Delta z)^4+5(\Delta z)^2\tau^2+6\tau^4\right)l_6+\cdots
\Big].
\label{eq.DssDps}
\end{eqnarray}
The leading term $2\Delta z \tau l_2$ contains
\begin{itemize}
\item
the familiar geometric ``vacuum'' contribution
$\Delta z \rho \tau \cos z\approx b\tau\cos z$; see Eqs.\ (\ref{eq.zforb})
and (\ref{eq.l2vac}),
\item
an atmospheric correction to $l_2$ of the order of 1--10 m/rad$^2$
(Fig.\ \ref{Ltayl.ps}).
It adds some tens of nanometers to the differential delay, if
$\tau <$ 1 arcmin $=3\cdot 10^{-4}$ rad, and $\Delta z < 3\cdot 10^{-5}$ rad
($b < 200$ m).
There is no equivalent contribution of this kind in planar earth models
like
Fig.\ \ref{DelModl.ps}.
\end{itemize}

\begin{figure}[h]
\plotone{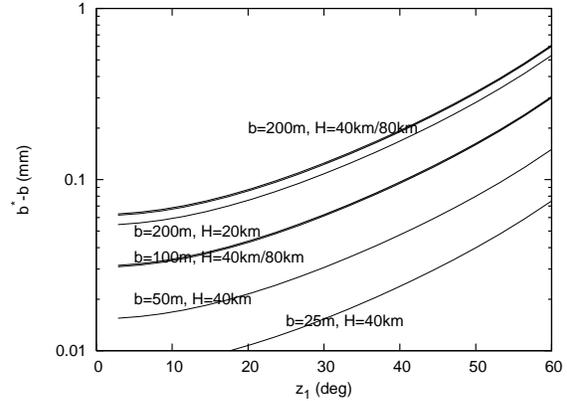}
\caption{
The difference $b^*-b$ between the length of the effective baseline
above the atmosphere and the geometric baseline on the ground according
to Eq.\ (\ref{eq.bstarlen}) as a function of $z_1$,
parametrized through the model (\ref{eq.chiofr})--(\ref{eq.chiofrPar}),
which is cut off at $H=20$, 40 or 80 km.
}
\label{Beff.ps}
\end{figure}

With Eqs.\ (\ref{eq.PofIs}) and (\ref{eq.rtayl}),
\begin{eqnarray}
I_1 &=& \rho n_0 \sin(z_1-R_1) \\
&\approx&\rho n_0 \sin\Big(\bar z -\xi_0+(1-\xi_1)\frac{\Delta z}{2}\nonumber \\
&&\qquad -\xi_2\frac{(\Delta z)^2}{4}-\xi_3\frac{(\Delta z)^3}{8}-\ldots\Big) , \label{eq.I1ofDz}\\
I_2 &=& \rho n_0 \sin(z_2-R_2) \\
&\approx&\rho n_0 \sin\Big(\bar z -\xi_0+(\xi_1-1)\frac{\Delta z}{2}-\xi_2\frac{(\Delta z)^2}{4}\nonumber\\
&&\qquad +\xi_3\frac{(\Delta z)^3}{8}-\ldots\Big), \label{eq.I2ofDz}
\end{eqnarray}
we may expand $P^2$ in a power series of $\Delta z$,
\begin{eqnarray}
P^2\!\!\!&\approx&\!\!\!\left\{\rho n_0 \cos \bar z_0(1-\xi_1)\right\}^2(\Delta z)^2 \nonumber \\
&& +\frac{1}{12} \rho^2 n_0^2\cos \bar z_0 (1-\xi_1) \nonumber \\
&& \times \left\{
[(\xi_1-1)^3-6\xi_3]\cos \bar z_0+6\xi_2[1-\xi_1]\sin \bar z_0
\right\} \nonumber \\
&&\times (\Delta z)^4 +\ldots
\label{eq.Pofdeltaz}
\end{eqnarray}
We do the same for $D^2$ via Eq.\ (\ref{eq.Dofl}), and eventually
combine these power series of $\Delta z$ at the r.h.s.\
of Eq.\ (\ref{eq.beff}),
\begin{eqnarray}
&&\!\! b^{*2}\approx\Big\{(1-\xi_1)^2\rho^2n_0^2\cos^2 \bar z_0+l_1^2\Big\}
(\Delta z)^2 \nonumber\\
&&+ \bigg\{\frac{\rho^2 n_0^2(1-\xi_1)\cos \bar z_0}{12}
\Big([(\xi_1-1)^3-6\xi_3]\cos \bar z_0 \nonumber \\
&&\qquad +6\xi_2[1-\xi_1]\sin \bar z_0\Big)
  +\frac{l_1l_3}{2}\bigg\} (\Delta z)^4 \nonumber\\
&&\quad+\ldots \label{eq.bstarlen}
\end{eqnarray}
which turns into Eq.\ (\ref{eq.bforz}) in the vacuum limit. Examples 
of this ``baseline magnification'' introduced by the atmosphere are
shown in Fig.\ \ref{Beff.ps}; the effect becomes larger if the transition
into the free space is smoothed by choosing a large cut-off height $H$.

\subsection{3D Geometry}\label{sec.zeni3D}
\subsubsection{Geographic Coordinates}
The previous section dealt with the case where the telescopes, the star and
the earth center are coplanar. In the general case, the direction of the star
and the direction of the second telescope do not share the same azimuth $A$.
Let the telescopes have geographical latitudes $\Phi_i$
and longitudes $\lambda_i$ in a geocentric spherical coordinate system:
\begin{equation}
{\bf r}_i=\rho \left(\begin{array}{c}
\cos \lambda_i \cos\Phi_i \\
\sin \lambda_i \cos\Phi_i \\
\sin\Phi_i \\
\end{array}\right), \quad (i=1,2).
\end{equation}
For the Very Large Telescope Interferometer in Northern Chile $\Phi\approx -0.4298$ rad
and $\lambda\approx -1.229$ rad, for instance.
The baseline angle in Eq.\ (\ref{eq.bofZ}) becomes
\begin{equation}
\cos Z = \sin\Phi_1\sin\Phi_2+\cos\Phi_1\cos\Phi_2\cos(\Delta\lambda),
\label{eq.b3d}
\end{equation}
where
$\Delta \lambda \equiv \lambda_1-\lambda_2$.
To transform Cartesian coordinates from the local alt-az-system of telescope $i$
(with the Cartesian coordinate $z$ pointing to the zenith, $x$ horizontally
tangentially to the earth toward north and the local horizon as indicated by
the dotted line in Fig.\ \ref{PointRho.ps}, $y$ horizontally toward west)
to the geocentric system (with $z$ pointing from the earth center to the north pole, $x$
from the center to the equator south of Greenwich, $y$ from the center
to the equator 1000 km west of Sumatra) we translate the coordinates
into a tilted system originating from the earth center, then (de)rotate them:
\begin{eqnarray}
{\bf r}_c&=&\left(
\begin{array}{ccc}
-\sin \Phi_i \cos \lambda_i & \sin\lambda_i & \cos\Phi_i\cos\lambda_i \\
-\sin \Phi_i \sin \lambda_i & -\cos\lambda_i & \cos\Phi_i\sin\lambda_i \\
\cos \Phi_i & 0 & \sin\Phi_i \\
\end{array}
\right) \nonumber \\
&&\quad \cdot \left ({\bf r}_i +\left( \begin{array}{c} 0 \\ 0 \\ \rho \end{array} \right) \right),
\quad i=1,2.
\label{eq.rgeo}
\end{eqnarray}
The inverse operation with the inverse matrix (which equals the 
transpose matrix) is
\begin{eqnarray}
{\bf r}_i\!\!\!&=&\!\!\!\left(
\begin{array}{ccc}
-\sin \Phi_i \cos \lambda_i & -\sin\Phi_i\sin\lambda_i & \cos\Phi_i \\
\sin \lambda_i & -\cos\lambda_i & 0 \\
\cos \Phi_i\cos\lambda_i & \cos\Phi_i\sin\lambda_i & \sin\Phi_i \\
\end{array}
\right) \cdot {\bf r}_c \nonumber \\
&&\quad -\left( \begin{array}{c} 0 \\ 0 \\ \rho \end{array} \right) .
\label{eq.rgeoinv}
\end{eqnarray}
The product of two such operations with indexes 1 and 2 converts the two
alt-az systems:
starting in Eq.\ (\ref{eq.rgeo}) with ${\bf r}_2=0$, computing ${\bf r}_c$,
and inserting this into (\ref{eq.rgeoinv}) for $i=1$ shows
that the origin of coordinates of telescope 2 is located at
\begin{eqnarray}
&&\!\!\!{\bf b}_{12}= \rho\cdot \nonumber \\
&&\!\!\!\left(
\begin{array}{c}
-\sin \Phi_1 \cos \Phi_2 \cos(\Delta\lambda)+\cos\Phi_1\sin\Phi_2 \\
\cos\Phi_2 \sin(\Delta\lambda) \\
\cos \Phi_1\cos\Phi_2 \cos(\Delta\lambda) +\sin\Phi_1\sin\Phi_2 -1\\
\end{array}
\right)
\nonumber
\end{eqnarray}
seen from the origin of telescope 1\@.
The length of this vector is $b=|{\bf b}_{12}|$
of Eq.\ (\ref{eq.bofZ}), using Eq.\ (\ref{eq.b3d});
the third coordinate is negative since the
second telescope lies below the horizon of the first telescope (and vice versa),
as illustrated in Fig.\ \ref{PointRho.ps}.
\subsubsection{Vacuum limit}
If the atmosphere is absent, the star direction is defined as
\begin{equation}
{\bf s}_i =\left(
\begin{array}{c}
\cos A_i \sin z_i \\
\sin A_i \sin z_i \\
\cos z_i \\
\end{array}
\right),\qquad i=1,2, \label{eq.stars}
\end{equation}
in terms of the true local azimuth $A_i$ and true zenith angle $z_i$ in
the $i$th telescope
coordinate system. $A$ is counted positive
starting from N to W\@. [This azimuth convention is the one
of \cite[\S II]{Smart}; the alternative convention of
\citep{Taff,Karttunen} is obtained with
the replacement $A_i\rightarrow \pi-A_i$.]
The cosine of the angle
between the star and the baseline in Fig.\ \ref{PointRho.ps} is
\begin{equation}
\cos \varphi_1 = {\bf s}_1 \cdot {\bf b}_{12}/b , \label{eq.blen}
\end{equation}
where ${\bf s}_1 \cdot {\bf b}_{12}$ is known as the geometric optical path delay.
If one swaps the indexes $1$ and $2$, the cosine switches its sign, because
in this parallax-free situation the angle between star and baseline is the
$180 ^\circ$-complement of the angle relative to the other telescope:
\begin{equation}
\cos \varphi_2 = {\bf s}_2 \cdot {\bf b}_{21}/b = - \cos\varphi_1.
\end{equation}
This may be verified with the standard coordinate transformations
between the hour angles $h_i$ and right ascension $\delta$
for $i=1,2$,
\begin{equation}
\cos\delta \sin h_i = \sin z_i \sin A_i,
\end{equation}
\begin{equation}
\sin z_i \cos A_i = \cos \Phi_i \sin\delta -\sin\Phi_i\cos\delta \cos h_i,
\label{eq.sinzcosA}
\end{equation}
\begin{equation}
\cos z_i = \sin \Phi_i \sin\delta +\cos\Phi_i\cos\delta \cos h_i,
\label{eq.zofRA} \label{eq.cosz}
\end{equation}
\begin{equation}
h_1-h_2 =\lambda_1-\lambda_2
\end{equation}
So if the atmosphere is absent, this angle
relates to $P=|\sin \varphi_i|b$ ($i=1,2$) as shown in Fig.\ \ref{Base.ps}.

The mean and difference in the true zenith angles remain defined as 
in Eqs.\ (\ref{eq.zbar}) and
\begin{equation}
\Delta z \equiv z_1-z_2,
\end{equation}
and can be retrieved from the geographical coordinates (\ref{eq.sinzcosA})--(\ref{eq.cosz})
and \cite[(4.3.34)-(4.3.37)]{AS}.

\subsubsection{Ray tracing}
\begin{figure}[hbt]
\plotone{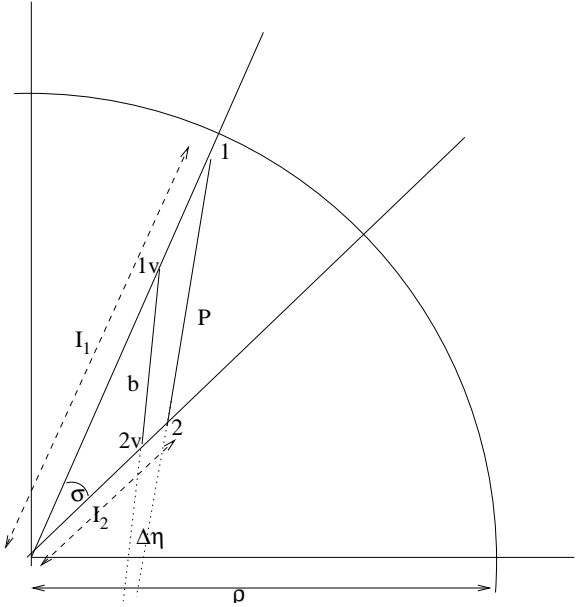}
\caption{A top view of the earth coordinates as seen from the incoming two rays
in which both ${\bf s}_i$ are perpendicular to the plane of the drawing.
If the atmosphere were absent, the first ray would see telescope 1 at
the position 1v. The atmospheric refraction pulls rays toward
the earth center, which means it has to
relocate its impact at the top of the atmosphere to the actual image
position 1 above the atmosphere
to end up at telescope 1 on the ground. The same effect for ray 2 means
the projected baseline vector $P$ above the atmosphere from 1 to 2 is
both longer and tilted by $\Delta \eta$ compared to the vacuum case from
1v to 2v.
}
\label{DelModl3.ps}
\end{figure}

There is one distinguished geocentric coordinate system for
the case of a single star, shown in Fig.\ \ref{DelModl3.ps}, in
which the direction vectors of the incoming light above the atmosphere
have only a component along the polar axis.
The side view of this geometry reduces to Fig.\ \ref{DelModl2.ps} if
the positions
1, 1v, 2 and 2v lie on the same projected straight line.
(In the following, ``projected'' means projected onto a plane
perpendicular to the ray propagation above the atmosphere.)
The strategy to transform the star position to the projected baseline
(vector) $P$ is (i) to calculate the triangle formed by the vacuum baseline
and the earth center seen by the incoming light, and (ii) stretch this
radially outward to include the effect of the two difference refraction
angles $R_i$:
the two true zenith angles $z_i$ for both telescopes
are assumed to be given via Eq.\ (\ref{eq.zofRA}). The impact parameters
$I_i$ of the rays
for the vacuum case are $\rho \sin z_i$ and they equal the lengths of the
(projected) station vectors
from the earth center to the points 1v and 2v in Fig.\ \ref{DelModl3.ps}\@.
The vacuum baseline in Fig.\ \ref{DelModl3.ps} stretches from
1v to 2v, which has the projected length $b|\sin \varphi_i|$ as written
down in Eq.\ (\ref{eq.blen}).
The projected baseline aperture angle $\sigma$ in the triangle with
side lengths
$\rho \sin z_i$ and $b|\sin \varphi_i|$ relates to the impact parameters
and projected vacuum baseline by planar trigonometry \cite[4.3.148]{AS},
\begin{equation}
\sin^2 z_1+\sin^2 z_2 = 2 \sin z_1 \sin z_2 \cos \sigma 
+ \frac{b^2}{\rho^2}\sin^2 \varphi_1.
\end{equation}
$\sigma$ is the projection of $Z$ defined in Eq.\ (\ref{eq.b3d}).
The equation mingles the true zenith angles from the two telescope's point
of view with the angles $\varphi_i$, which represent the star distance from the
baseline direction. One may reduce this to the geographic
coordinates by multiplying
${\bf r}_1+{\bf b}_{12}={\bf r}_2$ with ${\bf s}_1$ to get
\begin{eqnarray}
\cos\sigma&=&\frac{\cos Z-\cos z_1\cos z_2}{\sin z_1\sin z_2}\\
&=& 1-\frac{\cos(z_1-z_2)-\cos Z}{\sin z_1\sin z_2}.
\end{eqnarray}
$\cos\sigma$ and $\cos Z$ are close to unity in contemporary
optical interferometry,
so the actual implementation avoids the use of the cosine
in favor of the haversine,
\begin{eqnarray}
&&\hav\sigma=
\frac{\sin\frac{Z+\Delta z}{2} \sin\frac{Z-\Delta z}{2}}{\sin z_1\sin z_2} \\
&&=
\frac{Z^2-(\Delta z)^2}{4\sin^2\bar z}
-\left(\frac{1}{16}-\frac{1}{48}\sin^2\bar z\right)
 \left(\frac{\Delta z}{\sin\bar z}\right)^4 \nonumber \\
&&\, -\frac{1}{48}\left(\frac{Z}{\sin\bar z}\right)^4
+\frac{1}{16}\frac{Z^2 (\Delta z)^2}{\sin^4\bar z} \nonumber \\
&&\,
-\left(45-30\sin^2\bar z+2\sin^4\bar z\right)
  \frac{(\Delta z)^6}{2880 \sin^6\bar z} \nonumber \\
&&\, +\left(3-\sin^2 \bar z\right)\frac{(\Delta z)^4Z^2}{192 \sin^6\bar z}
-\frac{(\Delta z)^2Z^4}{192 \sin^4\bar z} \nonumber \\
&&\, +\frac{Z^6}{1440 \sin^2\bar z}+\ldots , \label{eq.havsigma}
\end{eqnarray}
to protect against cancellation of significant digits,
and returns from there to the sine, if needed,
\begin{eqnarray}
&&\sin\sigma = 2\sqrt{\hav \sigma}\sqrt{1-\hav\sigma} \\
&&= 2\sqrt{\hav \sigma}\left[1-\sum_{j=1}^\infty \frac{(2j-3)!!}{j!}
  \left(\frac{\hav \sigma}{2}\right)^j \right] \nonumber \\
&&=\sqrt{\hav\sigma}\left(2-\hav\sigma-\frac{\hav^2\sigma}{4}
-\frac{\hav^3\sigma}{8}
\cdots\right). \nonumber
\end{eqnarray}
The special coplanar case of Sec.\ \ref{sec.zeni2D} is included
as $\Delta z=Z$ and $\sigma=0$ or $\pi$.
(On some Linux systems, where the {\tt cosl} library function
does not support the full accuracy of the 96 bit {\tt long double}
number representation, it makes sense to switch to alternative
high-precision implementations of the cosine \citep{Schonfelder}.)
Two calculations for the actual impact parameters
$I_i=\rho n_0 \sin z_0^{(i)}$ starting from
the given $z_i$ would be done as in the preceding, ``aligned'' geometry of
Sec.\ \ref{sec.zeni2D}, and these inserted into
\begin{equation}
I_1^2+I_2^2 = 2 I_1 I_2 \cos \sigma + P^2
\end{equation}
to calculate the projected baseline $P$ and to generalize Eq.\ (\ref{eq.PofIs}).
$P^2=(I_1-I_2)^2+4I_1I_2\hav\sigma$ comprises the terms of
Eq.\ (\ref{eq.Pofdeltaz}) of the constrained geometry augmented by
\begin{equation}
4I_1I_2\hav \sigma
=\left(\rho n_0\sin \bar z_0\right)^2
\frac{Z^2-(\Delta z)^2}{\sin ^2\bar z}
+\ldots
\end{equation}
as read from Eq.\ (\ref{eq.havsigma}) in combination with
Eqs.\ (\ref{eq.I1ofDz})--(\ref{eq.I2ofDz}).

To calculate the path difference $D=L_1-L_2$ between the two rays above the
atmosphere, Eq.\ (\ref{eq.Dv}) remains valid, but in general a Taylor
expansion akin to (\ref{eq.taylDv}) does not exist,
because $P\ge I_i\sin\sigma$ ($i=1,2$) excludes small $P$ at arbitrary $\sigma$.
Eq.\ (\ref{eq.rtayl}) remains in use to expand the refraction angle in a
neighborhood of $\bar z$, and so do Eqs.\ (\ref{eq.Lofl})--(\ref{eq.Dofl})
that depend only on zenith distances but not on azimuths.

\subsubsection{Baseline Rotation}
\begin{figure}[h]
\plotone{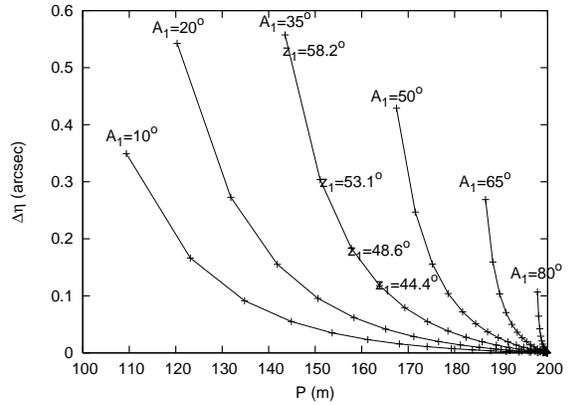}
\caption{The baseline rotation $\Delta \eta$ introduced in
Fig.\ \ref{DelModl3.ps} for a baseline $b=200$m, for 6 different azimuth
angles $A_1$ measured from $T_1$ toward the baseline, and for $\sin z_1$
changing in equidistant steps of $0.05$ from $0.05$ to $0.85$.
}
\label{ProtP.ps}
\end{figure}

\begin{figure}[h]
\plotone{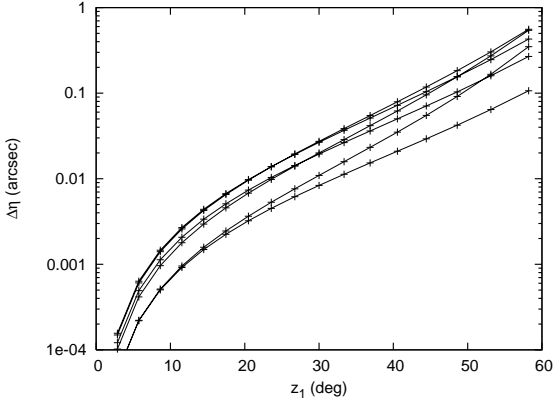}
\caption{An alternative view on the six lines of Fig.\ \ref{ProtP.ps} with the abscissa
switched from the projected baseline length to the azimuth angle.
Each small cross has a counterpart in Fig.\ \ref{ProtP.ps}. This graph would
not change visibly choosing a baseline of $b=100$ m.
}
\label{Protz.ps}
\end{figure}

The rotation angle between the baseline $b$ (projected on
a plane perpendicular to the star direction) and $P$ is $\Delta \eta\equiv\eta_P-\eta_b$
where $\eta_b$ and $\eta_P$ are the angles from $b$ to $I_2$ and
$P$ to $I_2$ respectively, and given by
\begin{eqnarray}
\frac{I_2}{I_1}&=&\cos\sigma +\sin \sigma \cot \eta_P, \\
\frac{\sin z_2}{\sin z_1}&=&\cos\sigma +\sin \sigma \cot \eta_b.
\end{eqnarray}
Numerical examples are presented in Figs.\ \ref{ProtP.ps}--\ref{Protz.ps}:
By symmetry, the rotation effect vanishes if the star azimuth is along
the baseline or perpendicular to it.
The angle $\Delta \eta$ may be about five times larger than the 
interferometric resolution of a 200 m baseline in the K-band:
the interferometric fringes appear slightly rotated on the detector,
and the true $(u,v)$ coordinate is found by rotating the ``apparent''
vector by $-\Delta\eta$.

There is no intrinsically new aspect compared to the analysis of Sec.\ \ref{sec.zeni2D}:
The rotation of the baseline vector $P$ relative to the projection of $b$
would again be absorbed into the pointing direction of the two telescopes.
With only one (scalar) observable, which is the differential delay, one cannot
measure both the astrometric angle (distance) between
two stars and the positioning angle of the S relative to the P
at the same time.

\subsubsection{Astrometric Case}
Eq.\ (\ref{eq.DssDps}) obviously looses its meaning
because the secondary star now has two degrees of freedom
and can no longer be positioned by a single angle $\tau$.
The details follow
by writing down Eq.\ (\ref{eq.cosz}) for both telescopes and both stars.
If the star distances $\tau_h$ and $\tau_\delta$ are defined as
\begin{equation}
h_{\mathrm S}\equiv h_{\mathrm P}+\tau_h,\qquad
\delta_{\mathrm S}\equiv \delta_{\mathrm P}+\tau_\delta,\qquad
\end{equation}
the expansion of Eq.\ (\ref{eq.cosz}) yields
for $z_{i\mathrm S}\equiv z_{i\mathrm P}+\tau_{iz}$
to lowest order in $\tau_h$ and $\tau_\delta$:
\begin{eqnarray}
\tau_{iz}&=&\frac{\cos\Phi_i \sin\delta_{\mathrm P}\cos h_{\mathrm P}
-\sin\Phi_i\cos\delta_{\mathrm P}}{\sin z_{i\mathrm P}}\tau_\delta
\nonumber \\
&& +\frac{\cos\Phi_i \cos\delta_{\mathrm P}\sin h_{\mathrm P}}
{\sin z_{i\mathrm P}}\tau_h+\ldots
\end{eqnarray}
The changes $\tau_{iA}$ in the azimuths are not detailed here, since
the refraction is determined by the zenith angles; they are
expected to ensure that the associated change in the star direction
(\ref{eq.stars}) forms an acute angle to the baseline to maintain the
interferometric resolution.
The symmetry suggested in Fig.\ \ref{Lofz.ps} will generally be broken:
$\Delta z_\textrm{P}\neq \Delta z_\textrm{S}$.
Restarting from Eq.\ (\ref{eq.Dofl}), the differential delay is expanded in
powers of the doubly differential $\Delta \tau_z$:
\begin{eqnarray}
&&D_\textrm{S}-D_\textrm{P} \nonumber \\
&& =\Delta z_{\mathrm P}\bar\tau_z \Big[
2 l_2
+3\bar\tau_z l_3 
+\left(
  (\Delta z_{\mathrm P})^2+ 4\bar\tau_z^2\right)l_4 \nonumber \\
&& \qquad\qquad +\cdots \Big]+O(\Delta \tau_z) ,
\end{eqnarray}
with $\bar \tau_z\equiv (\tau_{1z}+\tau_{2z})/2$ and
$\Delta \tau_z\equiv \tau_{1z}-\tau_{2z}$, where
the $l_j$ are again evaluated at the mean primary zenith,
$(z_{1\mathrm P}+z_{2\mathrm P})/2$.

\section{Summary}
The optical path length integral of star light passing through the
atmosphere can be handled with numerical and analytical methods known from
treatments of the more familiar refractivity integral. Subtraction of two
of these computes the optical path difference, and renormalizes the
optical path lengths to start from a common plane tangential to the
earth's upper atmosphere, which at the same time defines the projected baseline
in this plane perpendicular to the two rays. Definition of a right triangle
above the atmosphere with the projected baseline and the path difference
as two catheti defines a hypotenuse, which is an effective baseline that
is longer than and rotated relative to the geometric baseline between
the receiving telescopes on earth.

\appendix
\section{Atmospheric Refraction Models}\label{sec.nofr}
For the model (\ref{eq.chiofr}), the misassignment in the star position is
a few mas for each kilometer
of error in the scale height $K$.
In the limit
$K\rightarrow 0$, we recover the values for the flat earth of
Eq.\ (\ref{eq.Rofnflat});
the corresponding mathematical argumentation is given by
\cite[(4)]{StonePASP108}
and the limit $H^*\rightarrow 0$ in \cite[(1)]{Livengood}. 
In practice, the scale height is coupled to the atmospheric gas density,
and in a self-consistent model of a single 
average molecule species like Eq.\ (\ref{eq.chiofr}), $K$ is uniquely
coupled to the atmospheric pressure at ground level \citep{McCartney1976}.

In a variant of this problem, the scale heights of
various gas components differ and their mixing ratios change. In this case,
Eq.\ (\ref{eq.chiofr}) would be upgraded to a sum over gas components
with individual pairs of $\chi_0$ and $K$---it is doubtful to
scale the entire, single scale height with the ground humidity
instead \citep{Livengood}.
Consider for instance a fixed, ``dry air'' contribution to $\chi(r)$ with
a constant scale height of 10 km at $\chi_0=4.0808\cdot 10^{-4}$,
superimposed by water vapor at $\chi_0=1.75\cdot 10^{-6}$ that ``freezes out''
at scale heights of only a few kilometers:
the pointing variations are $\approx 25 \mu$as
per km change in the water vapor scale height.
Calibration of this water column could be achieved through
monitoring spectral ranges of high atmospheric water absorption
\citep{AkesonSPIE4006,MeisnerSPIE4838}.

The small difference in the refractive indexes at wavelengths of 2 and
2.4 $\mu$m, about
$\Delta\chi_0\approx 2.1\cdot 10^{-7}$,
displaces these two colored rays horizontally
by about 0.80 mm if they hit the earth surface, calculated at apparent zenith
angles of $z_0\approx 30$ deg.
If one looks at the starlight as an unvignetted plane wave, there is no such
effect.
Instead, there is a rainbow effect caused by this lensing of the earth
atmosphere, which spreads the apparent positions of these two colors
on the sky by $12.5$ mas,
which can be estimated through Eq.\ (\ref{eq.Rofnflat})
as $\Delta R\approx \tan z_0 \Delta n$ using
$\Delta n\approx \Delta \chi_0/2$ with Eq.\ (\ref{eq.chiofn}).
The calculation within the ray optics, however, can be used
to consider the distortion/decorrelation induced by turbulence on length
scales of these displacements \citep{ColavitaAO26_4113}.

\section{A Numerical Approach to the Refractivity Integrals}\label{sec.Rnum}
In the course of this investigation, the integrals (\ref{eq.RofnInt2})
have been decomposed into refraction within the open interval
$\rho \le r < \rho+H$, plus one kink at $r=\rho+H$ where $\epsilon=n^2$
changes abruptly from 1 to $\epsilon_{r=\rho+H}$:
\begin{eqnarray}
R_j&=&\frac{I^{2j+1}}{2}\int_1^{\epsilon_{r=\rho}} \frac{d\epsilon}{\epsilon(r^2\epsilon-I^2)^{j+1/2}} \\
&=&
\frac{I^{2j+1}}{2}\int_{\epsilon_{r=\rho+H}}^{\epsilon_{r=\rho}} \frac{d\epsilon}{\epsilon(r^2\epsilon-I^2)^{j+1/2}}
+
I^{2j+1}\int_1^{n_{r=\rho+H}} \frac{dn}{n[(\rho+H)^2n^2-I^2]^{j+1/2}}. \label{eq.Rkink}
\end{eqnarray}
The second term is Snell's law in terms of the two angles of incidence,
$\sin\psi_H\equiv I/(\rho+H)$ and
$n_{r=\rho+H}\sin\psi_{H-}\equiv \sin\psi_H$ just above
and below the atmosphere top boundary,
\begin{equation}
\int_1^{n_{r=\rho+H}} \frac{dn}{n(\frac{(\rho+H)^2n^2}{I^2}-1)^{j+1/2}}
=\int_{\psi_{H-}}^{\psi_H} \tan^{2j}x\, dx=
\left\{
\begin{array}{cl}
\psi_H-\psi_{H-} &, j=0 \\
\tan(\psi_H)-\psi_H-\ldots  &, j=1 \\
 \left[ \frac{\tan ^2 (\psi_H)}{3}-1 \right] \tan(\psi_H)+\psi_H
-\ldots  &, j=2,
\end{array}
\right.
\end{equation}
where the three dots mean the previous expression is to be
subtracted with all $\psi_H$ replaced by $\psi_{H-}$.
The first term in Eq.\ (\ref{eq.Rkink}) is mapped onto the interval $0\le t\le 1$
through the substitution
\begin{equation}
t=\gamma\frac{r-\rho}{r-\rho+H(1-\gamma)}.
\end{equation}
$\gamma$ is chosen with the idea of important sampling
such that $r_{1/2}$---typically selected as 2 km---is mapped on $t=1/2$.
This yields $\gamma=(H-r_{1/2})/(H-2r_{1/2})$ and
leaves
\begin{equation}
\int_{\epsilon_{r=\rho+H}}^{\epsilon_{r=\rho}} \frac{d\epsilon}{\epsilon(r^2\epsilon-I^2)^{j+1/2}}
=\gamma H(\gamma-1)\int_1^0 \frac{\frac{d\epsilon}{dr}dt} {(t-\gamma)^2 \epsilon (r^2\epsilon-I^2)^{j+1/2}} .
\end{equation}
This has been integrated with the trapezoidal rule, doubling the number $N$ of equidistant
sampling
points after each step. As $\epsilon (r^2\epsilon-I^2)^{j+1/2}$ is a smooth function
over the $t$-interval, the error in the trapezoidal rule is dominated by the
variation in $(d\chi/dr)/(t-\gamma)^2$. For the exponential
model (\ref{eq.chiofr}) one can demonstrate
that the sequence $V_N$, $V_{2N}$, $V_{4N}$,\ldots,
obtained by repeated division of the sampling step size by two, obeys
$V_{4N}-V_{2N}\approx\frac{1}{2}(V_{2N}-V_N)$. The Richardson extrapolation
induces the estimator
$V \stackrel{N\to \infty}{\longrightarrow} 2V_{2N}-V_N$,
{\em not\/} the Simpson rule;
monitoring the convergence of this estimator has been used to terminate
the subdivision loop.
A very similar analysis is applicable to the integrals $v_i$.

\bibliography{eso,all}

\end{document}